\documentclass[journal]{IEEEtran}

\usepackage{comment}
\usepackage{ifpdf}
\usepackage{filecontents,lipsum}
\usepackage[noadjust]{cite}
\usepackage{cite}
\usepackage[pdftex]{graphicx}
\usepackage{amsmath}
\usepackage{algorithmic}
\usepackage{array}
\usepackage[caption=false,font=normalsize,labelfont=sf,textfont=sf]{subfig}
\usepackage{fixltx2e}
\usepackage{stfloats}
\usepackage{url}
\usepackage[dvipsnames]{xcolor}

\begin{document}

\title{Fast Spectrogram Inversion using Multi-head Convolutional Neural Networks}

\author{
  Sercan \"{O}. Ar{\i}k$^{*}$,
  Heewoo Jun$^{*}$,
  Gregory Diamos

\thanks{Baidu Silicon Valley Artificial Intelligence Lab 1195 Bordeaux Dr. Sunnyvale, CA 94089.}
\thanks{$^{*}$Equal contribution}
\thanks{Manuscript received August, 2018.}
}

%
%

\markboth{ }%
{Shell \MakeLowercase{\textit{et al.}}: Bare Demo of IEEEtran.cls for IEEE Journals}

%


\maketitle
\begin{abstract}
We propose the multi-head convolutional neural network (MCNN) for waveform synthesis from spectrograms. Nonlinear interpolation in MCNN is employed with transposed convolution layers in parallel heads. MCNN enables significantly better utilization of modern multi-core processors than commonly-used iterative algorithms like Griffin-Lim, and yields very fast (more than 300x real-time) runtime. For training of MCNN, we use a large-scale speech recognition dataset and losses defined on waveforms that are related to perceptual audio quality. We demonstrate that MCNN constitutes a very promising approach for high-quality speech synthesis, without any iterative algorithms or autoregression in computations. 

\end{abstract}

\begin{IEEEkeywords}
Phase reconstruction, deep learning, convolutional neural networks, short-time Fourier transform, spectrogram, time-frequency signal processing, speech synthesis.
\end{IEEEkeywords}

\IEEEpeerreviewmaketitle

\vspace{-0.5 cm}
\section{Introduction}
\vspace{-0.1 cm}

A spectrogram contains intensity information of time-varying spectrum of a waveform. Waveform to spectrogram conversion is fundamentally lossy, because the magnitude calculation removes the phase from the short-time Fourier transform (STFT). Spectrogram inversion has been studied widely in literature. Yet, there is no known algorithm that guarantees 
a globally optimal solution at a low
computational complexity. A fundamental challenge is the non-convexity of intensity constraints with an unknown phase. 

The most popular technique for spectrogram inversion is the Griffin-Lim (GL) algorithm \cite{GriffinLim}. GL is based on iteratively estimating the unknown phases by repeatedly converting between frequency and time domain using the STFT and its inverse, substituting the magnitude of each frequency component to the predicted magnitude at each step. Although the simplicity of GL is appealing, it can be slow due to the sequentiality of operations. In \cite{fastGL}, a fast variant is studied by modifying its update step with a term that depends on the magnitude of the previous update step. In \cite{SPSI}, the single-pass spectrogram inversion (SPSI) algorithm is introduced, which can synthesize waveforms in a single fully deterministic pass and can be further improved with extra GL iterations. SPSI estimates the instantaneous frequency of each frame by peak-picking and quadratic interpolation. In \cite{noniterativePhase_prusa}, another non-iterative spectrogram inversion technique is proposed, based on the partial derivatives with respect to a Gaussian window, which allows analytical derivations. In \cite{convexGL}, a convex relaxation is applied to express spectrogram inversion as a semidefinite program with a convergence guarantee, at the expense of the increased dimensionality. Overall, one common drawback for these generic spectrogram inversion techniques is their fixed objectives, rendering them inflexible to adapt for a particular domain like human speech. 

One common use case of spectrograms is the audio domain, which is also the focus of this paper. Autoregressive modeling of waveforms, in particular for audio, is a common approach. State-of-the-art results in generative speech modeling use neural networks \cite{parallelwavenet}\cite{shen2017natural} that employ autoregression at the sample rate. Yet, these models bring challenges for deployment, as they need to run inference $\sim$16k-24k times every second. One approach is to approximate autoregression with an inference-efficient model which can be trained by learning an inverse-autoregressive flow using distillation \cite{parallelwavenet}. Recently, autoregressive neural networks have also been adapted for spectrogram inversion. \cite{DeepVoice2} uses the WaveNet architecture \cite{oord:2016:wavenet}, which is composed of stacked dilated convolution layers with spectrogram frames as external conditioner. But autoregression at sample rate is employed, resulting in slow synthesis. A fundamental question is whether high quality synthesis necessitates explicit autoregressive modeling. Some generative models, e.g. \cite{DeepVoice3}, \cite{Tacotron1}, synthesize audio by applying autoregression at the rate of spectrogram timeframes (100s of samples), and still does not yield a noticeable decrease in audio quality. 

We propose the multi-head convolutional neural network (MCNN) that employs non-autoregressive modeling for the perennial spectrogram inversion problem. Our study is mainly motivated by two trends. Firstly, modern multi-core processors, such as GPUs or TPUs \cite{TPU_datacenter}, achieve their peak performance for algorithms with high compute intensity \cite{gpu_compute}. Compute intensity (also known as operational intensity) is defined as the average number of operations per data access. Secondly, many recent generative audio models, such as text-to-speech \cite{DeepVoice3}\cite{Tacotron1}, audio style transfer \cite{audiostyletransfer}, or speech enhancement \cite{donahue2017segan}, output spectrograms (that are typically converted to waveforms using GL), and can potentially benefit from direct waveform synthesis by integrating trainable models into their end-to-end frameworks. MCNN achieves very high audio quality (quantified by human raters and conventional metrics like spectral convergence (SC) and speaker classification accuracy), while achieving more than 300x real-time synthesis, and has the potential to be integrated with end-to-end training in audio processing.

\vspace{-0.1cm}

\section{Multi-head Convolutional Neural Network}

We assume the STFT-magnitude input
for the waveform $s$, $|\textrm{STFT}(s)|$, has a dimension of $T_{spec} \times F_{spec}$ and the corresponding waveform has a dimension of $T_{wave}$, where $T_{spec}$ is the number of spectrogram timeframes, $F_{spec}$ is the number of frequency channels, and $T_{wave}$ is the number of waveform samples. The ratio $T_{wave}/T_{spec}$ is determined by the spectrogram parameters, the hop length and the window length. We assume these parameters are known a priori.

\vspace{-0.4cm}

\begin{figure}[!ht]
    \centering
    \includegraphics[width=0.43\textwidth]{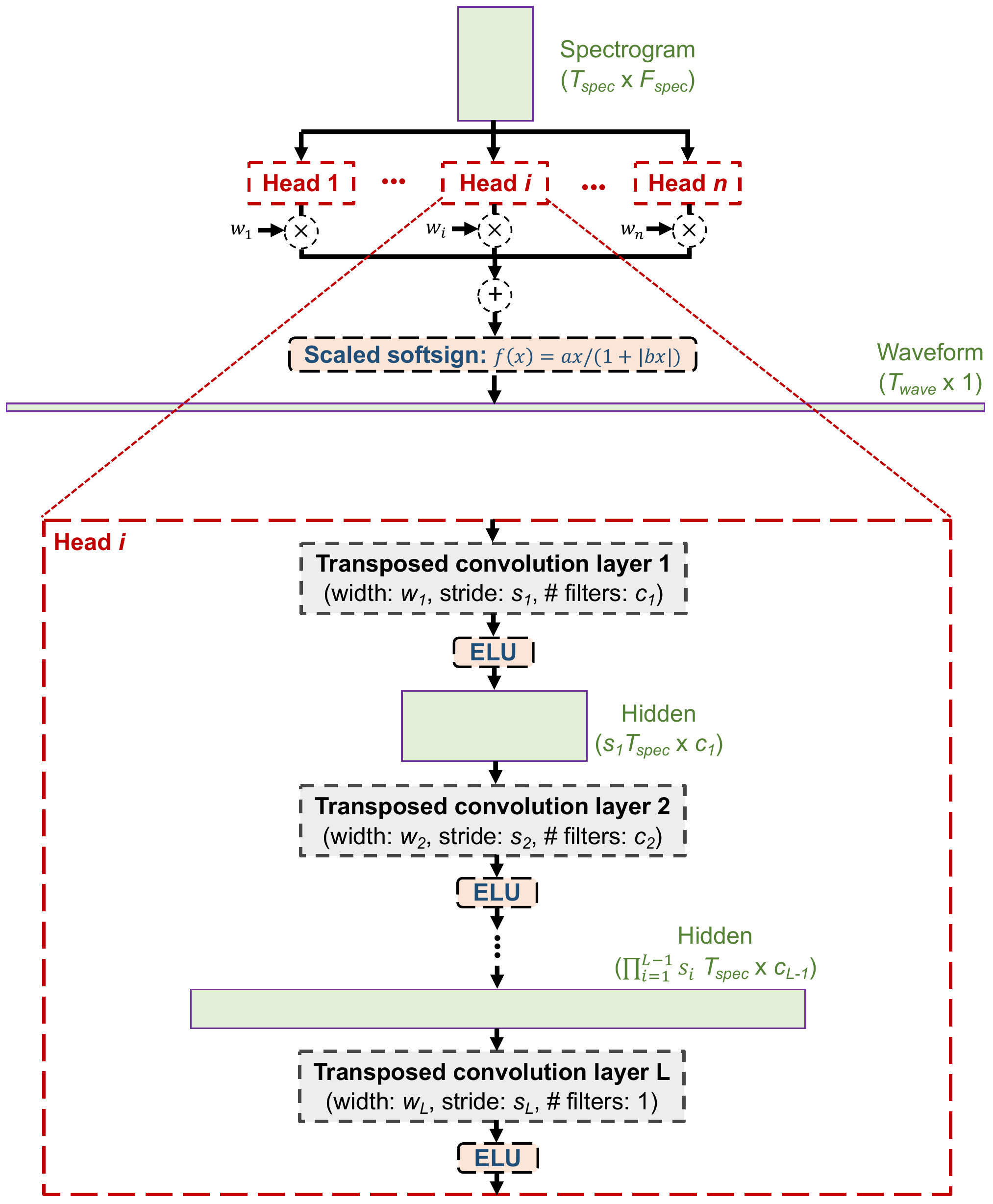}
    \vspace{-0.2cm}
    \caption{Proposed MCNN architecture for spectrogram inversion.}
    \label{fig:gen_model}
\end{figure}
\vspace{-0.2cm}

To synthesize a waveform from the spectrogram, a function parameterized by a neural network needs to perform nonlinear upsampling in time domain, while utilizing the spectral information in different channels. Typically, the window length is much longer than the hop length, and it is important to utilize this extra information in neighboring time frames. For fast inference, we need a neural network architecture that achieves a high compute intensity by repeatedly applied computations with the same kernel. 

Based on these motivations, we propose the multi-head convolutional neural network (MCNN) architecture. MCNN has multiple heads that use the same types of layers but with different weights and initialization, and they learn to cooperate as a form of ensemble learning. By using multiple heads, we allow each model to allocate different upsampling kernels to different components of the waveform which is analyzed further in Appendix \ref{multihead_analysis}. Each head is composed of $L$ transposed convolution layers (please see \cite{tansposeconv} for more details about transposed convolutional layers), as shown in Fig. \ref{fig:gen_model}. Each transposed convolution layer consists of a 1-D temporal convolution operation, followed by an exponential linear unit \cite{clevert2016elu}\footnote{It was empirically found to produce superior audio quality than other nonlinearities we tried, such as ReLU and softsign.}. For the $l^\text{th}$ layer, $w_l$ is the filter width, $s_l$ is the stride, and $c_l$ is the number of output filters (channels). Striding in convolutions determines the amount of temporal upsampling, and should be chosen to satisfy $\prod_{l=1}^{L} s_l \cdot T_{spec}= T_{wave}$. Filter widths control the amount of local neighborhood information used while upsampling. The number of filters determine the number of frequency channels in the processed representation, and should be gradually reduced to 1 to produce the time-domain waveform. As the convolutional filters are shared in channel dimension for different timesteps, MCNN can input a spectrogram with an arbitrary duration.
A trainable scalar is multiplied to the output of each head to match the overall scaling of inverse STFT operation and to determine the relative weights of different heads.
Lastly, all head outputs are summed and passed through a scaled softsign nonlinearity, $f(x) = ax/(1 + |bx|)$, where $a$ and $b$ are trainable scalars, to bound the output waveform.

\vspace{-0.2cm}
\section{Audio Losses}
\label{audio_losses}

Loss functions that are correlated with the perceptual quality should be used to train generative models. We consider a linear combination of the below loss terms
between the estimated waveform $\hat{s}$ and the ground truth waveform $s$,
presented in the order of observed empirical significance:
\\
\textbf{(i) Spectral convergence (SC):}
\begin{equation}
\label{SC_eq}
    {\| |\textrm{STFT}(s)| - |\textrm{STFT}(\hat{s})| \|_{F}} / {\||\textrm{STFT}(s)|\|_{F}},
\end{equation}
where $\|\cdot\|_{F}$ is the Frobenius norm over time and frequency. SC loss emphasizes highly on large spectral components, which helps especially in early phases of training.\\
\textbf{(ii) Log-scale STFT-magnitude loss:}
\begin{equation}
\label{log_STFT_eq}
    \|\log (|\textrm{STFT}(s)| + \epsilon) - 
    \log(|\textrm{STFT}(\hat{s})| + \epsilon)\|_1,
\end{equation}
where $\|\cdot\|_1$ is the $L^1$ norm and $\epsilon$ is a small number. The goal with log-scale STFT-magnitude loss is to accurately fit small-amplitude components (as opposed to the SC), which tends to be more important towards the later phases of training.\\
\textbf{(iii) Instantaneous frequency loss:}
\begin{equation}
\label{IF_eq}
    \left \| \frac{\partial }{\partial t} \phi ({\textrm{STFT}}(s)) - 
    \frac{\partial }{\partial t} \phi ( {\textrm{STFT}}(\hat{s})) \right \|_1,
\end{equation}
where $\phi(\cdot)$ is the phase argument function. The time derivative $\frac{\partial }{\partial t}$ is estimated with finite difference $\frac{\partial f}{\partial t} = \frac{f(t + \Delta t) - f(t)}{ \Delta t}$. Spectral phase is highly unstructured along either time or frequency domain, so fitting raw phase values is very challenging and does not improve training. Instead, instantaneous frequency is a smooth phase-dependent metric, which can be more accurately fit. \\
\textbf{(iv) Weighted phase loss:}
\begin{align}
\| & | \textrm{STFT}(s) | \odot | \textrm{STFT} (\hat{s})| 
 - \Re \{ \textrm{STFT}(s) \} \odot \Re \{ \textrm{STFT} (\hat{s})\} 
\notag \\
& - \Im \{ \textrm{STFT}(s) \} \odot \Im \{ \textrm{STFT} (\hat{s})\} 
\|_1,
\label{weighted_phase_eq}
\end{align} 
where $\odot$ is element-wise product, $\Re$ is the real part and $\Im$ is the imaginary part. When a circular normal distribution is assumed for the phase, the log-likelihood function is proportional to $L(s, \hat{s}) = \cos(\phi ({\textrm{STFT}}(s)) - \phi ({\textrm{STFT}}(\hat{s})))$ \cite{WavenetAutoencoder}. We can correspondingly define a loss as $W(s, \hat{s}) = 1 - L(s, \hat{s})$, which is minimized ($W(s, \hat{s}) = 0$) when $\phi ({\textrm{STFT}}(s)) = \phi ({\textrm{STFT}}(\hat{s}))$. To focus on the high-amplitude components more and for better numerical stability, we further modify $W(s, \hat{s})$ by scaling it with $|\textrm{STFT}(s) | \odot | \textrm{STFT}(\hat{s})|$, which yields Eq. \ref{weighted_phase_eq} after $L^1$ norm. 

\vspace{-0.4cm}
\section{Experimental Results}

\subsection{Experimental setup}

We use the LibriSpeech dataset~\cite{panayotov2015librispeech}, after a preprocessing pipeline, composed of segmentation and denoising, similar to \cite{DeepVoice3}. LibriSpeech contains 960 hours of public-domain audiobooks from 2484 speakers sampled at 16 KHz. It is originally constructed for automatic speech recognition and the audio quality is thus lower compared to speech synthesis datasets. 

As the spectrogram parameters, a hop length of 256 (16 ms duration), a Hanning window with a length of 1024 (64 ms duration), and an FFT size of 2048 are assumed. MCNN has 8 transposed convolution layers, with $(s_i, w_i, c_i) = (2, 13, 2^{8-i})$ for $1 \leq i \leq 8$, i.e. halving in the number of channels is balanced with temporal upsampling by a factor of two. The coefficients of the loss functions in Sec. \ref{audio_losses} are chosen as 1, 6, 10 and 1 respectively, optimized for the audio quality by employing a random grid search. The model is trained using the Adam optimizer \cite{ADAM}. The initial learning rate of 0.0005 is annealed at a rate of 0.94 every 5000 iterations. The model is trained for $\sim$600k iterations with a batch size of 16 distributed across 4 GPUs with synchronous updates. We compare our results to conventional implementations of GL \cite{GriffinLim} and SPSI \cite{SPSI} with and without extra GL iterations. 

\vspace{-0.4cm}
\subsection{Synthesized audio waveform quality}

\begin{figure}[!ht]
    \centering
    \includegraphics[width=0.49\textwidth]{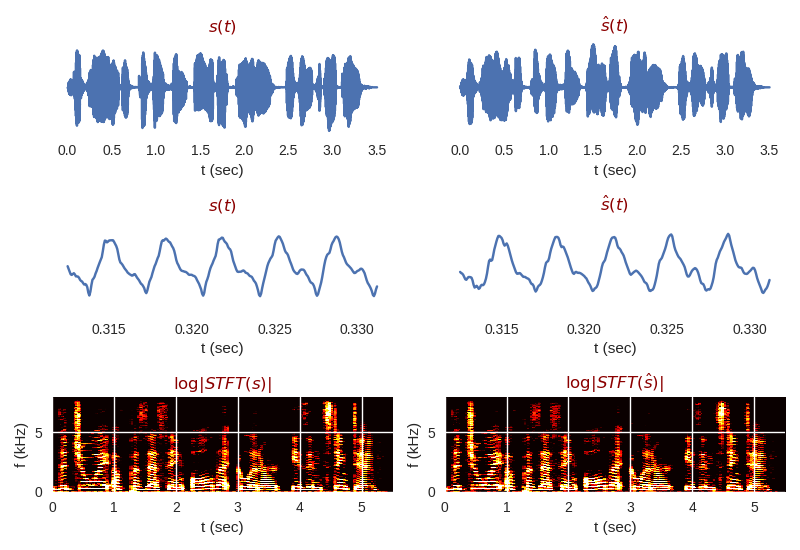}
    \vspace{-0.3cm}
    \caption{Comparison of the waveform (entire utterance and a zoomed portion) and its spectrogram, for the ground truth (left) and MCNN-generated (right).}
    \label{example1}
\end{figure}

A synthesized audio waveform is exemplified in Fig. \ref{example1}. We observe that complicated patterns can be fit, and there is a small phase error between relevant high-amplitude spectral components (the amount of shift between the peaks is low).

We evaluate the quality of synthesis on the held-out LibriSpeech samples (Table \ref{tab:librispeech_results}) using mean opinion score (MOS)\footnote{Human ratings are collected via Amazon Mechanical Turk framework independently for each evaluation, as in \cite{voicecloning}. Multiple votes on the same sample are aggregated by a majority voting rule.}, SC, and classification accuracy (we use the speaker classifier model from \cite{voicecloning}) to measure the distinguishability of 2484 speakers.\footnote{Audio samples can be found in \url{https://mcnnaudiodemos.github.io/}.} 

\begin{figure}[!ht]
    \centering
    \includegraphics[width=0.4\textwidth]{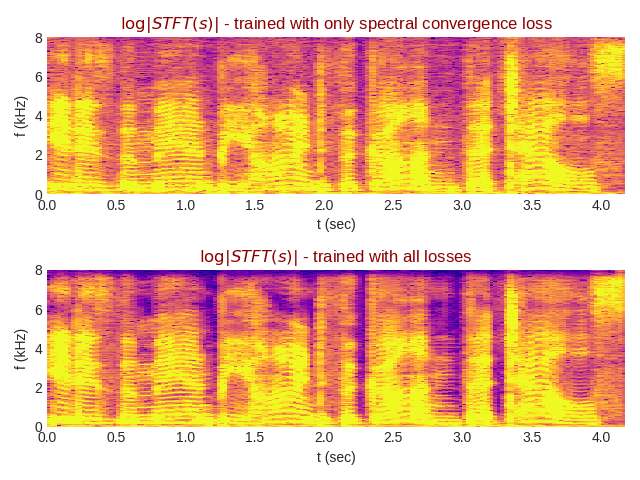}
    \vspace{-0.3cm}
    \caption{Log-STFT of synthesized sample for MCNN trained with only SC loss (top) and all losses (bottom).}
    \label{spect1}
\end{figure}

According to the subjective human ratings (MOS), MCNN outperforms GL, even with a high number of iterations and SPSI initialization. When trained only on spectral convergence (SC), MCNN is on par with GL. Indeed, merely having SC loss as the training objective yields even slightly better SC for test samples. Yet, with only SC loss, lower audio quality is observed for some samples due to generated background noise and less clear high frequency harmonics, as exemplified in Fig. \ref{spect1}. To further improve the audio quality, flexibility of MCNN for integration of other losses is beneficial, as seen from Table \ref{tab:librispeech_results}. Ablation studies also show sufficiently large filter width and sufficiently high number of heads are important. Transposed convolutions tend to produce checkerboard-like patterns \cite{odena:2016:deconvolution}, and a single-head may not be able to generate all frequencies efficiently. In an ensemble, however, different heads cooperate to cancel out artifacts and cover different frequency bands, as further elaborated in Appendix \ref{multihead_analysis}. Lastly, high speaker classification accuracy shows that MCNN can efficiently preserve the characteristics of speakers (e.g. pitch, accent, etc.) without any conditioning, showing potential for direct integration into training for applications like voice cloning.

\begin{table*}[ht!]
	\centering
	\caption{MOS with 95\% confidence interval, average spectral convergence and speaker classification accuracy for LibriSpeech test samples.}
	\small\addtolength{\tabcolsep}{-5pt}
		\begin{tabular}{|c|c|c|c|}
		\hline
		Model & MOS (out of 5) & Spectral convergence (dB) & Classification accuracy (\%) \\ \hline\hline
 		\textbf{MCNN (filter width of 13, 8 heads, all losses)}  & 3.50 $\pm$ 0.18 & $-12.9$ &  76.8 \\ \hline
		MCNN (filter width of 9) & 3.26 $\pm$ 0.18 & $-11.9$ & 73.2 \\ \hline
	   	MCNN (2 heads) & 2.78 $\pm$ 0.17 & $-10.7$ & 71.4 \\ \hline
	   	MCNN (loss: Eq. (\ref{SC_eq})) & 3.32 $\pm$ 0.16 & $-13.3$ &  69.6 \\ \hline
	   	MCNN (loss: Eq. (\ref{SC_eq}) \& Eq. (\ref{log_STFT_eq})) & 3.35 $\pm$ 0.18 & $-12.6$ & 73.2 \\ \hline
        GL (3 iterations)               & 2.55 $\pm$ 0.26 & $-5.9$ & 76.8 \\\hline
 		GL (50 iterations)              & 3.28 $\pm$ 0.24 & $-10.1$ & 78.6 \\ \hline
  		GL (150 iterations)              & 3.41 $\pm$ 0.21 & $-13.6$ & 82.1 \\ \hline
		SPSI                            & 2.52 $\pm$ 0.28 & $-4.9$ & 75.0 \\ \hline
		SPSI + GL (3 iterations)        & 3.18 $\pm$ 0.23 & $-8.7$ & 78.6 \\ \hline
		SPSI + GL (50 iterations)       & 3.41 $\pm$ 0.19 & $-11.8$ & 78.6  \\ \hline
		Ground truth  & 4.20 $\pm$ 0.16  & $-\infty$ & 85.7 \\ \hline
		\end{tabular}
	\label{tab:librispeech_results} 
\end{table*}
\vspace{-0.2cm}

\subsection{Generalization and optimization to a particular speaker}

The audio quality is maintained even when the MCNN trained on LibriSpeech is used for an unseen speaker (from a high-quality text-to-speech dataset \cite{DeepVoice1}), as shown in Table \ref{tab:beijing_results}. To evaluate how much the quality can be improved, we also train a separate MCNN model using only that particular speaker's audio data, with reoptimized hyperparameters.\footnote{Filter width is increased to 19 to improve the resolution for modeling of more clear high frequency components. Lower learning rate and more aggressive annealing are applied due to the small size of the dataset, which is $\sim$20 hours in total. Loss coefficient of Eq. \ref{log_STFT_eq} is increased because the dataset is higher in quality and yields lower SC.} The single-speaker MCNN model yields a very small quality gap with the ground truth.

\begin{table}[ht!]
	\centering
	\caption{MOS with 95\% confidence interval for single-speaker samples (from an internal dataset \cite{DeepVoice1}).}
	\small\addtolength{\tabcolsep}{-5pt}
		\begin{tabular}{|c|c|c|c|}
		\hline
		Model & MOS (out of 5) \\ \hline\hline
 		\textbf{MCNN (trained on LibriSpeech)}  & 3.55 $\pm$ 0.17      \\\hline
 		\textbf{MCNN (trained on single-speaker}  & 3.91 $\pm$ 0.17 \\\hline
 		GL (150 iterations)              & 3.84 $\pm$ 0.16  \\ \hline
		SPSI + GL (50 iterations)        & 3.69 $\pm$ 0.17 \\ \hline
		Ground truth            & 4.28 $\pm$ 0.14   \\ \hline
		\end{tabular}
	\label{tab:beijing_results}
\end{table}

\vspace{-0.2cm}
\subsection{Representation learning of the frequency basis}

\begin{figure}[!ht]
    \centering
    \includegraphics[width=0.45\textwidth]{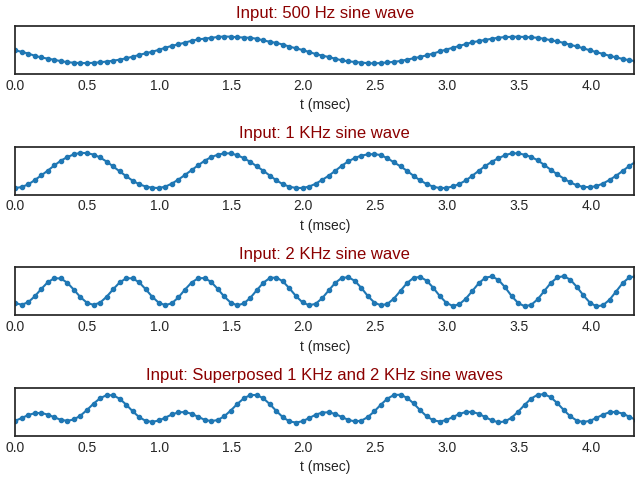}
    \vspace{-0.3cm}
    \caption{Synthesized waveforms by MCNN (trained on LibriSpeech), for spectrogram inputs corresponding to sinusoids at 500, 1000 and 2000 Hz, and for a spectrogram input of superposed sinusoids at 1000 and 2000 Hz.}
    \label{sinusoid}
\end{figure}

MCNN is trained only with human speech, which is composed of time-varying signals at many frequencies. Interestingly, MCNN learns the Fourier basis representation in the spectral range of human speech, as shown in Fig. \ref{sinusoid} (representations get poorer for higher frequencies beyond human speech, due to the increased train-test mismatch). When the input spectrograms correspond to constant frequencies, sinusoidal waveforms at those frequencies are synthesized. When the input spectrograms correspond to a few frequency bands, the synthesized waveforms are superpositions of pure sinusoids of constituent frequencies. For all cases, phase coherence over a long time window is observed.

\subsection{Deployment considerations}

We evaluate the inference complexity and compute intensity (based on the assumptions presented in Appendix \ref{complexity_modeling}) and benchmark runtime on a Nvidia Tesla P100 GPU.\footnote{We consider the Tensorflow implementation of operations without specific kernel optimizations, which can yield to further improvements specific to the hardware. For a fair comparison, we consider the GPU implementation of GL using Tensorflow FFT/inverse FFT operations.} The baseline MCNN model from Table 1 (the one in bold font) can generate $\sim$5.2M samples/sec, yielding $\sim$330 times faster-than-real-time waveform synthesis. Compared to MCNN, the runtime of GL is $\sim$20 times slower for 50 iterations, and $\sim$60 times slower for 150 iterations. The computational complexity of MCNN is $\sim$2.2 GFLOPs/sec, and indeed is slightly higher than the complexity of 150 GL iterations. However, much shorter runtime is due to the properties of the neural network architecture that render it very well suited for modern multi-core processors like GPUs or TPUs. First and foremost, MCNN requires much less DRAM bandwidth (in byte/s) - the compute intensity of MCNN, 61 FLOPs/byte, is more than an order of magnitude higher than that of GL, 1.9 FLOPs/byte. In addition, MCNN has a shorter critical path of dependent operations in its compute graph compared to GL, yielding parallelization and utilization. Efficient inference with such a highly-specialized model is enabled by learning from large-scale training data, which is not possible for signal processing algorithms like GL.

\section{Conclusions}

We propose the MCNN architecture for the spectrogram inversion problem. MCNN achieves very fast waveform synthesis without noticeably sacrificing the perceptual quality. MCNN is trained on a large-scale speech dataset and can generalize well to unseen speech or speakers. MCNN and its variants will benefit even more from future hardware in ways that autoregressive neural network models and traditional iterative signal processing techniques like GL cannot take advantage of. In addition, they will benefit from larger scale audio datasets, which are expected to close the gap in quality with ground truth. An important future direction is to integrate MCNN into end-to-end training of other generative audio models, such as text-to-speech or audio style transfer systems.

\newpage

\bibliographystyle{IEEEtran}
\bibliography{bibliography}

\newpage
\appendix

\subsection{Complexity modeling}
\label{complexity_modeling}

Computational complexity of operations is represented by the total number of algorithmic FLOPs without considering hardware-specific logic-level implementations. (Such a complexity metric also has limitations of representing some major sources of power consumption, such as loading and storing data.) We count all point-wise operations (including nonlinearities) as 1 FLOP, which is motivated with the trend of implementing most mathematical operations as a single instruction. We ignore the complexities of register memory-move operations. We assume that a matrix-matrix multiply, between $W$, an $m \times n$ matrix and $X$, an $n \times p$ matrix takes $2mnp$ FLOPs. Similar expression is generalized for multi-dimensional tensors, that are used in convolutional layers. For real-valued fast Fourier transform (FFT), we assume the complexity of $2.5 N \log_2(N)$ FLOPs for a vector of length $N$ \cite{FFTW}. For most operations used in this paper, Tensorflow profiling tool \cite{TFprof} includes FLOP counts, which we directly adapted. 

\subsection{Analysis of contributions of multiple heads}
\label{multihead_analysis}

\begin{figure*}[!ht]
    \centering
    \includegraphics[width=0.85\textwidth]{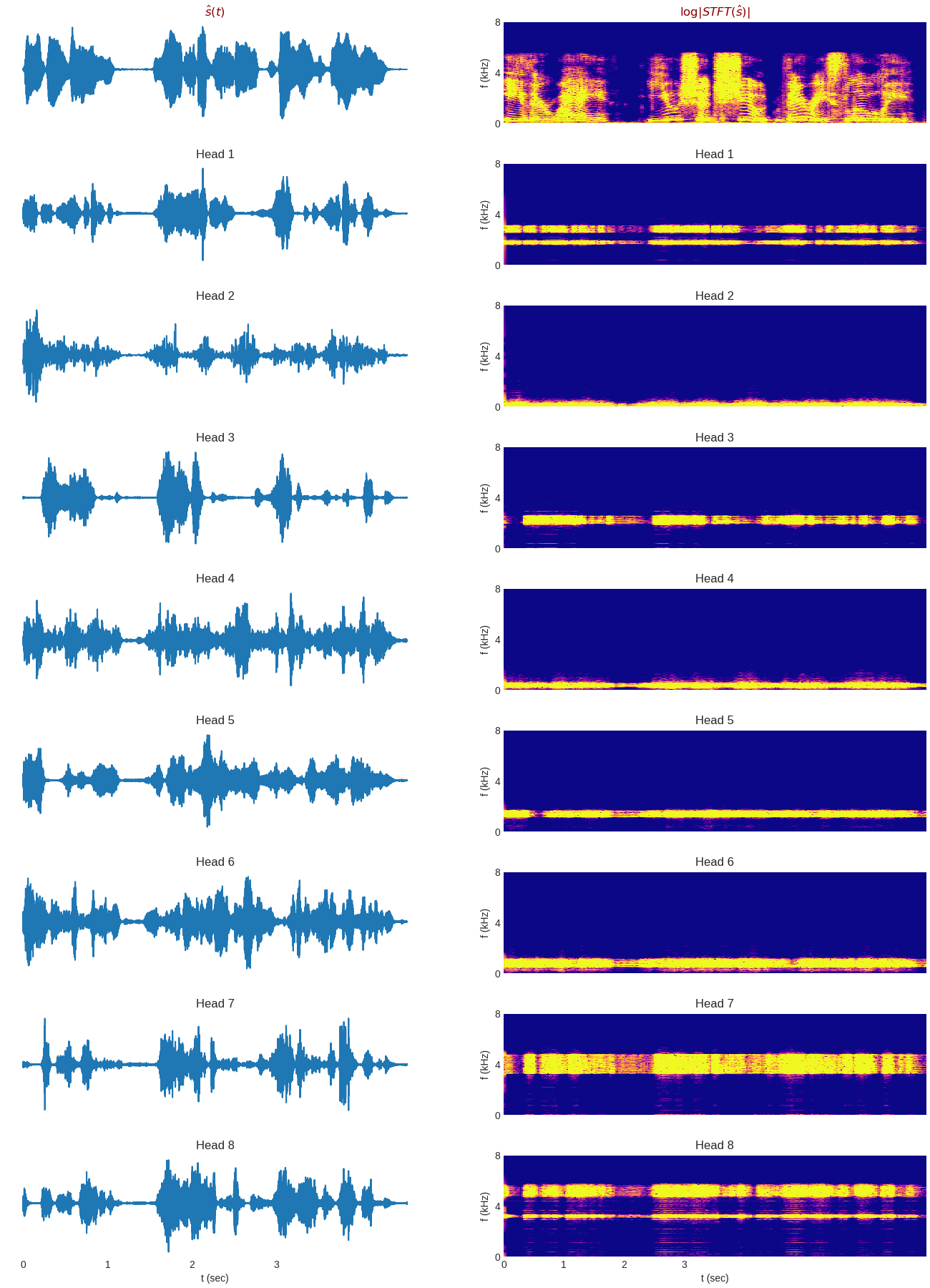}
    \caption{Top row: An example synthesized waveform and its log-STFT. Bottom 8 rows: Outputs of the waveforms of each of the constituent heads. For better visualization, waveforms are normalized in each head and small-amplitude components in STFTs are discarded after applying a threshold.}
    \label{multihead}
\end{figure*}

Fig. \ref{multihead} shows the outputs of individual heads along with the overall waveform. We observe that multiple heads focus on different portions of the waveform in time, and also on different frequency bands. For example, head 2 mostly focuses on low-frequency components. While training, individual heads are not constrained for such a behavior. In fact, different heads share the same architecture, 
but initial random weights of the heads determine which portions of the waveform they will focus on in the later phases of training. The structure of the network promotes cooperation with the end-to-end objective. Hence, initialization with the same weights would nullify the benefit of the multi-head architecture.
Although intelligibility of individual waveform outputs is very low (we also note that a nonlinear combination of these waveforms can also generate new frequencies that do not exist in these individual outputs.), their combination can yield highly natural-sounding waveforms.

\end{document}